\documentclass[aps,prl,showpacs,floatfix,superscriptaddress,twocolumn,color]{revtex4}
\usepackage{graphicx}
\usepackage{amsmath}
\usepackage{amsfonts}
\usepackage{amssymb}
\usepackage{wasysym}
\usepackage{dsfont}
\usepackage{color}
\usepackage{soul}
\usepackage{natbib}
\usepackage{doi}

\definecolor{redmarker}{rgb}{0.9,0.0,0.0}
\definecolor{greenmarker}{rgb}{0.0,0.5,0.0}
\definecolor{bluemarker}{rgb}{0.1,0.1,0.9}
\definecolor{violetmarker}{rgb}{0.6,0.1,0.99}
\definecolor{luismarker}{rgb}{1.0,0.5,0.00}

\renewcommand{\Re}{\operatorname{Re}}
\renewcommand{\Im}{\operatorname{Im}}
\newcommand{\Tr}{\operatorname{Tr}}

\begin{document}


\title{{Synchronization in the presence of distributed delays}}


\author{Lucas Wetzel}		
\affiliation{Max Planck Institute for the Physics of Complex Systems, N\"othnitzer Str. 38, 01187 Dresden, Germany}
\author{Luis G. Morelli}		
\affiliation{Max Planck Institute for the Physics of Complex Systems, N\"othnitzer Str. 38, 01187 Dresden, Germany}
\affiliation{Max~Planck~Institute~of~Molecular~Cell~Biology~and~Genetics,~Pfotenhauerstr.~108,~01307~Dresden,~Germany}
\affiliation{IBioBA, Max Planck Society Partner Institute, Godoy Cruz 2390, C1425FQD, Buenos Aires, Argentina}%
\affiliation{Departamento de F\'{\i}sica, FCEyN UBA, Ciudad Universitaria, 1428 Buenos Aires, Argentina}%
\author{Andrew C. Oates}		
\affiliation{Max~Planck~Institute~of~Molecular~Cell~Biology~and~Genetics,~Pfotenhauerstr.~108,~01307~Dresden,~Germany}
\affiliation{Department of Cell and Developmental Biology, University~College~London,~Gower~Street,~London~WC1E~6BT,~UK}
\affiliation{Francis Crick Institute, Mill Hill Laboratory, The Ridgeway, London NW7 1AA, UK}
\author{Frank J\"ulicher}		 \email{julicher@pks.mpg.de}
\affiliation{Max Planck Institute for the Physics of Complex Systems, N\"othnitzer Str. 38, 01187 Dresden, Germany}
\author{Sa\'ul Ares}			\email{saul@math.uc3m.es}
\affiliation{Max Planck Institute for the Physics of Complex Systems, N\"othnitzer Str. 38, 01187 Dresden, Germany}
\affiliation{Grupo Interdisciplinar de Sistemas Complejos (GISC), and Departamento de Matem\'aticas, Universidad Carlos III de Madrid, 28911 Legan\'es, Spain}

\date{\today}

\begin{abstract}  
\noindent 
We study systems of identical coupled oscillators introducing a distribution of delay times in the coupling.
For arbitrary network topologies, we show that the frequency and stability of the fully synchronized states depend only on the mean of the delay distribution. 
However, synchronization dynamics is sensitive to the shape of the distribution. 
In the presence of coupling delays, the synchronization rate can be maximal for a specific value of the coupling strength.
\end{abstract}
\pacs{05.45.Xt, 
02.30.Ks, 
87.10.-e
}

\maketitle


In complex systems, dynamic states arise from the interaction of many subunits. 
Time delays in these interactions, for example due to finite communication times, can have a profound impact on collective dynamics \cite{atay10}.
In systems of coupled oscillators, time delays in the coupling can affect the collective frequency as well as
synchronization behavior \cite{SchusterWagner89,niebur91,Yeung99,zanette00,jeong02,Earl03,Acebron05,montbrio06,sethia08,sethia10,eguiluz11}.
Time delays with a unique, well defined value are often called discrete delays.
It has been shown that full synchronization of oscillators can be achieved in the presence of discrete time delays in the coupling~\cite{SchusterWagner89}.
Interestingly multiple synchronized states can exist for the same value of the delay~\cite{Yeung99}.
Exact criteria for the stability of these synchronized states have been derived~\cite{Earl03,papachristodoulou05,papachristodoulou06}.

Discrete coupling delays are the simplest way to introduce interactions that are not instantaneous. 
However, in many situations it is important to consider more realistic distributed coupling delays \cite{cooke82,Macdonald2008,bressloff99,atay03,omi08,kyrychko11,laing11,skardal12,kyrychko12}.
This is the case when different past times affect the present state with different weights \cite{megerle08,munsky09,Marquez-Lago2010}.
Coupled oscillators with delayed coupling play an important role for a wide variety of systems in physics, chemistry, biology and engineering \cite{SynchRosenPikovKurths,manrubia,Earl03,Morelli2009,Herrgen2010,AresMorelli2012}.
Examples are the synchronization of
electronic circuits \cite{srinivasan11},
lasers \cite{wunsche05,franz08},
the flashing of large populations of fireflies \cite{smith35,tyrrel08},
the coordination of many cellular oscillators in a tissue \cite{Lewis2003,ay2013} and mobile devices in networks \cite{papachristodoulou05,papachristodoulou06}.
Examples of systems where distributed delays are relevant include the study of social dynamics \cite{iribarren09}, neuronal networks \cite{roberts08,atay06}, 
ecology \cite{eurich05}, epidemiology \cite{jin06,mccluskey10} or genetic oscillations \cite{feng10}.

In this letter we study synchronization in systems of oscillators with memory kernels in the coupling that account for distributed time delays.
These kernels may describe the annealed limit of a system with noisy delays, 
in contrast to the quenched limit with discrete heterogeneous delays~\cite{papachristodoulou05,LeeOttAntonsen09}.
%
%
%
%
%
We show that the stability of synchronized states 
does not depend on the shape of the distribution function describing delay times.
In contrast, the relaxation time to the synchronized state does depend on the shape of this distribution function and is important for the synchronization process.
Furthermore, synchronization can be optimized for a particular value of the coupling strength in the presence of coupling delays.

We consider a systems of coupled oscillators with distributed coupling delays
\begin{equation}
   \frac{d \theta_k(t)}{dt}=\omega+\frac{K}{n_k}\sum_{l=1}^{N}c_{kl}h\left(\int_{0}^{\infty} ds\, g(s) \theta_l(t-s) -\theta_k(t)\right),
  \label{eqn:general-model}
\end{equation}
where $\theta_k(t)$ is the state of the $k$-th oscillator, $\omega$ is the intrinsic frequency of individual oscillators, $K$ is the coupling strength,
$n_k$ is the number of coupling connections for oscillator $k$, $N$ is the total number of oscillators in the system, $h(\theta)$ is a $2\pi$-periodic coupling function and
$g(s)$ denotes the delay distribution.
We consider normalized delay distributions fulfilling $\int_{0}^{\infty} ds\, g(s) = 1$, with mean $\bar\tau \equiv \int_{0}^{\infty} ds\, s\, g(s) $.
The coefficients $c_{kl}$ define the connectivity of the network: $c_{kl}=1$ if oscillator $k$ is connected to oscillator $l$ and $c_{kl}=0$ otherwise. 
We consider the case where the network does not consist of unconnected subnetworks and all oscillators receive at least one coupling signal.
This implies an absolute generality of network topologies included in our theory, which extends that of previous studies \cite{Earl03}.

The model in Eq.~(\ref{eqn:general-model}) is not gauge invariant, i.e., not invariant under the transformation $\theta_k \rightarrow \theta_k+2\pi$.
Therefore the variables $\theta_k$ are not phases.
The result of the integral in Eq.~(\ref{eqn:general-model}) depends on the definition of the variable $\theta_k$: for instance, for $\theta_k\,\in\,\left[0,2\pi\right)$ we would obtain different 
results than for $\theta_k\,\in\,\left(-\infty,\infty\right)$. 
%
%
We can analyze this problem making a change of variable to $X_k = e^{i\theta_k}$ \cite{jorg2014}. 
Using the Fourier series of the coupling function
\begin{equation}
h(\phi) = \sum_{m=-\infty}^\infty f_m e^{i m \phi},
\label{eq.fourier}
\end{equation}
we can write Eq.~(\ref{eqn:general-model}) as:
\begin{align}
\label{eqn:general-model-x}
&\frac{dX_k(t)}{dt}  = i \omega X_k(t) + \\
&i X_k(t) \frac{K}{n_k}\sum\limits_{l}c_{kl} \sum_m f_m e^{i\, m \int\limits_0^{\infty}ds\,g(s)\, \log X_l(t-s) } {X_k^*}^m(t). \nonumber
\end{align}
The gauge dependence appears now via the definition of the complex logarithm, which is a multivalued function.
To obtain an unambiguous expression $\log X_l$, one branch has to be chosen, which is equivalent to choosing a (2$\pi$-periodic) gauge for the phase.
%
For different branches of $\log X_l$, the value of the convolution with $g(s)$ is different. 
%
%
One could also interpret the complex logarithm in a different way: instead of choosing a branch, we can understand $\log X_l$ as a Riemann surface that covers the punctured ($X_l=0$ is excluded) complex plane in an infinite-to-1 way. 
%
Such a choice for $\log X_l$ is multivalued with no branch cuts, and once an initial value for $\log X_l$ is defined, it is continuous as long as $X_l$ is continuous.
%
%
This is equivalent to define the variable $\theta_k \in (-\infty, \infty)$ corresponding to a continuous ``phase variable'' that counts the winding number about $X_l=0$, which is what we do from here onwards.
%
%
For simplicity we keep using the terminology of phases in the following, keeping in mind that the $\theta_k$ are not true phase variables.\\

\begin{figure}[bt]
  \begin{center}
    \includegraphics[width=8.5cm]{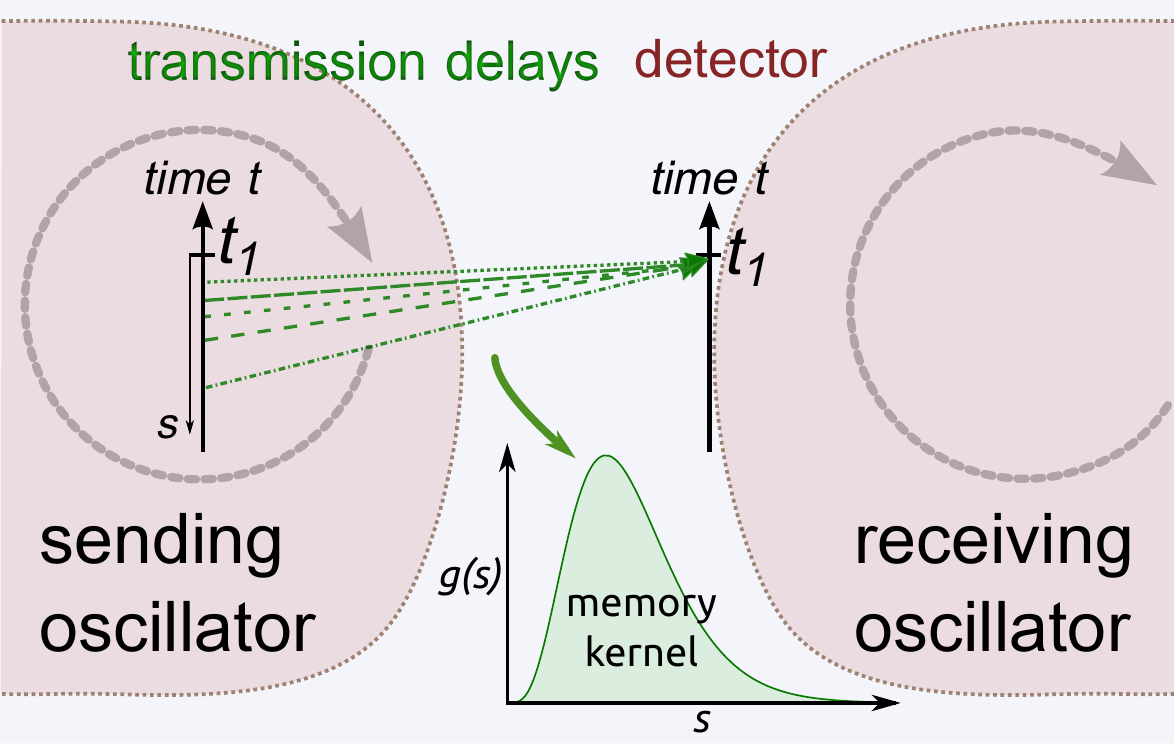}
\caption{Schematic representation of two oscillators coupled with distributed delays.
Transmission delays account for signals that originate at different past times $t_1-s$ from the sending oscillator and arrive at time $t_1$ at the detector of the receiving oscillator.
The distribution of transmission delay times is given by the the memory kernel $g(s)$.}
    \label{fig:SketchModel}
  \end{center}
\end{figure}


\noindent
\emph{Collective frequency of phase-locked states.}
Synchronized solutions are the most striking manifestation of dynamical order \cite{SynchRosenPikovKurths,manrubia}.
We look for phase-locked synchronized solutions of Eq.~(\ref{eqn:general-model}) with no phase lags: 
\begin{equation}
  \theta_k(t)=\Omega t,
  \label{eqn:synch-solution}
\end{equation}
where the phases of all oscillators are equal and grow linearly in time with a collective frequency $\Omega$.
We substitute this ansatz into Eq.~(\ref{eqn:general-model}) and obtain a transcendental equation for $\Omega$ \cite{SchusterWagner89,Yeung99,Earl03,papachristodoulou05,papachristodoulou06}:
\begin{equation} 
  \Omega=\omega+K h(-\Omega\bar{\tau}).
  \label{eqn:collective-freq-coherent}
\end{equation}
Eq.~(\ref{eqn:collective-freq-coherent}) is  independent of the number of oscillators in the system and the network topology. 
Furthermore, it depends only on the mean delay $\bar\tau$ and not on the particular shape of the delay distribution $g(s)$. 
This means that the functional dependence of the collective frequency of the synchronized state is identical for discrete delays $\tau$ and distributed delays with mean $\bar{\tau}=\tau$.\\

\noindent
\emph{Linear dynamics close to the synchronized state.}
Close to a synchronized state described by Eq.~(\ref{eqn:collective-freq-coherent}), the linearized dynamics of the system is studied by considering a weak perturbation $q_k(t)$ to Eq.~(\ref{eqn:synch-solution}):
\begin{equation}
  \theta_k(t)=\Omega t + \epsilon q_k(t),
  \label{eqn:synch-solution-perturb}
\end{equation}
with $\epsilon \ll 1$, and substituting into Eq.~(\ref{eqn:general-model}).
The linear dynamic equations for the perturbation are:
\begin{equation}
  \dot{q}_k(t)=\frac{\alpha}{n_k}\sum_{l=1}^{N} c_{kl} \left[\int_{0}^{\infty} ds\,g(s) q_l(t-s)-q_k(t)\right], 
  \label{eqn:perturb-dynamics}
\end{equation}
where
\begin{equation}
\alpha \equiv K h'(-\Omega \bar{\tau}). 
\end{equation}
We search for eigenmodes of the form $q_k(t)=v_k e^{\lambda t}$. 
If all values of $\Re(\lambda)$ are negative, the perturbation decays and the synchronized state is stable.
Using Eq.~(\ref{eqn:perturb-dynamics}) we obtain the characteristic equation:
\begin{equation}
  v_k\lambda=\frac{\alpha}{n_k}\sum_{l=1}^{N} c_{kl}\left[ v_l  \hat{g}(\lambda) - v_k \right],
  \label{eqn:charact-eq11}
\end{equation} 
where $\hat{g}(\lambda)$ is the Laplace transform of the delay distribution \cite{Widder1946,NIST2010}:
\begin{equation}
 \hat{g}(\lambda)\equiv \int_{0}^{\infty}ds\,g(s)e^{-\lambda s}.
\label{eq:ghat}
\end{equation}
The absolute value $\left|\hat{g}(\lambda)\right|$ has an upper bound independently of the shape of the distribution:
\begin{equation}
  \left| \hat{g}(\lambda) \right|  \leq 1,
    \label{eqn:laplace_absvalue}
\end{equation}
if $\Re(\lambda)\geq0$.
This property will be important to derive a general stability criterion for the synchronized states.

For $\alpha = 0$ the solution to Eq.~(\ref{eqn:charact-eq11}) is $\lambda = 0$ and the synchronized state is neutrally stable.
For $\alpha \neq 0$ and $\hat{g}(\lambda)^{-1}\neq 0$, Eq.~(\ref{eqn:charact-eq11}) can be rearranged and expressed as an eigenvalue problem:
\begin{equation}
  \zeta v_k = \sum_{l=1}^{N} d_{kl} v_l.
  \label{eqn:charact-eq-final1_matrix}
\end{equation}
The relation $n_k=\sum_{l=1}^{N}c_{kl}$ has been used, and $d$ is the normalized connectivity matrix with components $d_{kl} \equiv {c_{kl}}/{n_k}$.
These properties together with Gerschgorin's circle theorem \cite{Strang86,Earl03} imply for the eigenvalues $\zeta$ of the matrix $d$:
\begin{equation}
 {\left| \zeta \right|}\leq 1.
  \label{eqn:follow-bound}
\end{equation}
These eigenvalues relate to the values of $\lambda$ corresponding to the characteristic eigenmodes of the system by:
\begin{equation}
  \zeta = {\hat{g}(\lambda)}^{-1} \left( {\lambda}/{\alpha}+1 \right).
  \label{eqn:charact-eq-coherent-definitionzeta}
\end{equation}
%
This is the characteristic equation for the complex synchronization rates $\lambda$.
Combined with Eq.~(\ref{eqn:charact-eq-final1_matrix}), one can see that the eigenvector $\vec{v}=(1,\,1,\,\dots,\,1)^T$ with eigenvalue $\zeta=1$, implying $\lambda=0$, is always a solution. 
It corresponds to a neutrally stable mode reflecting the symmetry of Eq.~(\ref{eqn:general-model}) under a uniform phase shift of all oscillators. 
We exclude this trivial mode from our discussion.\\

\noindent
\emph{Stability of the synchronized state.}
Here we show that only the mean delay $\bar{\tau}$ of the delay distribution is relevant to the linear stability of the synchronized states.
Hence, stability is independent of other factors such as the shape of the delay distribution and network topology.
This result generalizes previous work to arbitrary delay distributions and general network topologies \cite{Earl03,papachristodoulou05,papachristodoulou06}. 
We find that synchronized states are stable (the largest non-trivial $\Re(\lambda)$ is negative) if and only if 
\begin{equation}
  \label{eq:criterion}
\alpha=Kh'(-\Omega \bar{\tau})>0.
\end{equation}
This stability criterion can be derived from Eq.~(\ref{eqn:charact-eq-coherent-definitionzeta}) as follows. We rewrite Eq.~(\ref{eqn:charact-eq-coherent-definitionzeta}) as
\begin{eqnarray}
  \alpha \left| \hat{g}(\lambda) \right| {\left| \zeta \right|} \cos(\psi+\xi) &=& \Re(\lambda)+\alpha, \label{eqn:eigenvalues_real} \\
  \alpha \left| \hat{g}(\lambda) \right| {\left| \zeta \right|} \sin(\psi+\xi) &=& \Im(\lambda),
  \label{eqn:eigenvalues_imag}
\end{eqnarray}
where we express the complex numbers $\hat{g}(\lambda)$ and $\zeta$ by their magnitudes and phases:
\begin{eqnarray}
  \hat{g}(\lambda) &\equiv&\left| \hat{g}(\lambda) \right|e^{i\psi},  \\
  \zeta &\equiv&\left| \zeta \right|e^{i\xi}.  
\end{eqnarray}
%
%
%

Using Eq.~(\ref{eqn:laplace_absvalue}), Eq.~(\ref{eqn:follow-bound}), Eq.~(\ref{eqn:eigenvalues_real}) and Eq.~(\ref{eqn:eigenvalues_imag}), we can now prove the stability criterion, Eq.~(\ref{eq:criterion}).
First we assume there exists a $\lambda$ satisfying Eq.~(\ref{eqn:charact-eq-coherent-definitionzeta}), such that $\Re(\lambda)\geq0$ for $\alpha>0$ and show that this leads to a contradiction.
%
From Eqs.~(\ref{eqn:eigenvalues_real}-\ref{eqn:eigenvalues_imag}), for $\alpha=\left| \alpha \right|$ we obtain:
\begin{equation}
  \left| \hat{g}(\lambda) \right|^2 {\left| \zeta \right|}^2 = 1+{(\Re(\lambda)^2+\Im(\lambda)^2+2\left|\alpha \right|\left|\Re(\lambda)\right|)}/{ \alpha^2}.
  \label{eqn:stabproof1}
\end{equation}
Since $\left| \zeta \right| \leq 1$ and $\left| \hat{g}(\lambda) \right| \leq 1$ for all $\Re(\lambda) \geq 0$,
it follows that $\left| \hat{g}(\lambda) \right|^2 {\left| \zeta \right|}^2\leq 1$.
For $\lambda \neq 0$ the right hand side of Eq.~(\ref{eqn:stabproof1}) is greater than $1$, which contradicts the assumption.
It thus follows that for $\alpha>0$ there are no solutions with $\Re(\lambda)\geq 0$.
Hence $\alpha>0$ assures $\Re(\lambda)< 0$ and is sufficient for the asymptotic stability of the synchronized states given by Eq.~(\ref{eqn:collective-freq-coherent}).

We now show that if $\alpha<0$, the synchronized state is either unstable or neutrally stable.
For negative $\alpha$ with $\alpha=-\left| \alpha \right|$, Eq.~(\ref{eqn:eigenvalues_real}) can be rewritten as: 
\begin{equation}
  -\left|\alpha\right| \left| \hat{g}(\lambda) \right| {\left| \zeta \right|} \, \cos(\psi+\xi) = \Re(\lambda)-\left|\alpha\right|.
  \label{eqn:stabproof2}
\end{equation}
%
%
For unknown $\psi$ and $\xi$, two cases have to be distinguished.
Case i: If $\cos(\psi+\xi)\leq 0$ we have:
\begin{equation}
  \Re(\lambda)=\left|\alpha\right| \left(1 + \left| \hat{g}(\lambda) \right| {\left| \zeta \right|} \, \left| \cos(\psi+\xi) \right|\right),
  \label{eqn:stabproof3}
\end{equation}
and it follows that $\Re(\lambda)> 0$.
Case ii: If $\cos(\psi+\xi)> 0$ we can write:
\begin{equation}
  \Re(\lambda)-\left| \alpha\right| = - \left|\alpha\right|  \left| \hat{g}(\lambda) \right| {\left| \zeta \right|} \, \left| \cos(\psi+\xi) \right|.
  \label{eqn:stabproof4}
\end{equation}
The sign of $\Re(\lambda)$ satisfying Eq.~(\ref{eqn:stabproof4}) is less obvious, but we can show that there are always non-trivial perturbation modes with $\Re(\lambda)\geq 0$.
The function $f(\lambda)= - \left|\alpha\right|  \left| \hat{g}(\lambda) \right| {\left| \zeta \right|} \, \left| \cos(\psi+\xi) \right|$ has the property $0\geq f(\Re(\lambda)=0)\geq - \left|\alpha\right| $
and $f(\lambda)\rightarrow0$ for $\Re(\lambda)\rightarrow\infty$.
The continuity of $f(\lambda)$ then requires that a value of $\lambda$ with $\Re(\lambda)\geq 0$ exists.
%
%
The network topology assures that non-trivial modes always exist,
and we have just shown that these modes cannot be asymptotically stable if $\alpha<0$.
%
Non-trivial modes are assured because the sum of all eigenvalues $\zeta$ is equal to the trace of matrix $d$. 
 Due to the connectedness of the network, not all diagonal elements of $d$ can be $1$, which implies $\Tr(d)<N$. 
 Since $d$ has $N$ eigenvalues, this means that not all $\zeta$ can be 1, assuring the existence of non-trivial modes.
We have also shown that for $\alpha>0$ these modes are always asymptotically stable,
so Eq.~(\ref{eq:criterion}) is the necessary and sufficient condition for the linear stability of the synchronized states given by Eq.~(\ref{eqn:collective-freq-coherent}). \\

\begin{figure}[t]
  \begin{center}
  \includegraphics[width=6.2cm]{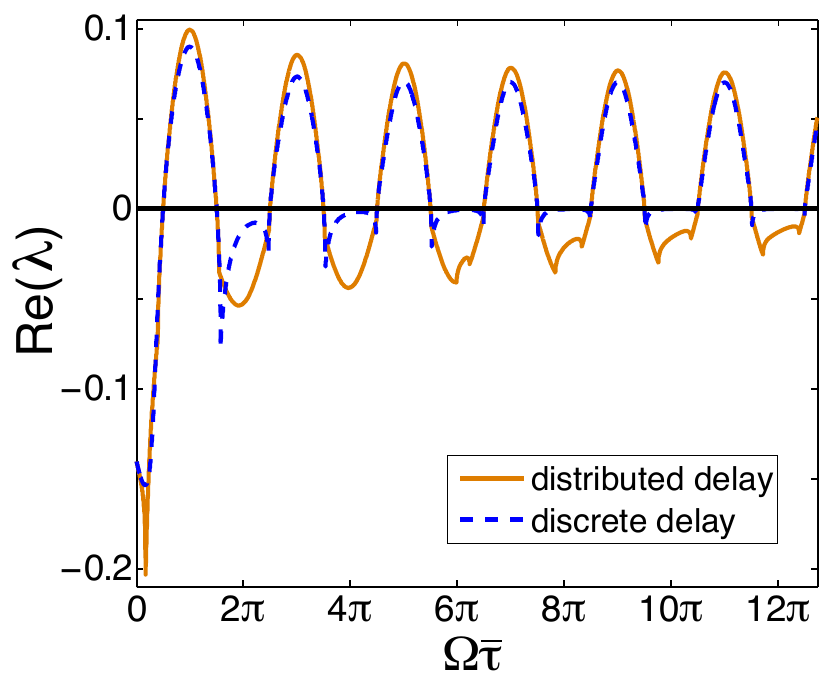}                                                    
  \caption{Stability, given by the sign of $\Re(\lambda)$, is independent of the delay distribution, but synchronization rate, proportional to $-\Re(\lambda)$, is not.
		  $\Re\left( \lambda \right)$ versus $\Omega\bar{\tau}$ for the eigenvalue $\lambda$ with the largest real part. Dashed curve: discrete delay. 
		  Continuous curve: distributed delay with  an exponential distribution of mean $\bar{\tau}$, $g(s)=e^{-s/\bar{\tau}}/\bar{\tau}$.
		  $N=2$, $n=1$, $\omega=0.223 \,\text{min}^{-1}$, $K=0.07 \,\text{min}^{-1}$, $h(\theta)=\sin(\theta)$.}
  \label{fig:perturb_decay_nearest_N2}
  \end{center}
\end{figure}

\noindent
\emph{Transient dynamics close to the synchronized state.}
The results presented above concern only limit cycles and provide no information on transient dynamics.
Transients are important because much can be learned about the dynamics of resynchronization from a perturbed synchronous state~\cite{RiedelKruse2007}.
Transient dynamics close to synchrony is characterized by the synchronization rate, which is proportional to $-\Re(\lambda)$. Positive $\Re(\lambda)$ means that perturbations grow and synchrony is unstable.
Inspection of Eq.~(\ref{eqn:charact-eq-coherent-definitionzeta}) suggests that for different delay distributions the synchronization rate might differ, namely,
the modes resulting from perturbations of the synchronized state can decay or grow with different rates for delay distributions with different shapes.
This is indeed the case: we show in Fig.~\ref{fig:perturb_decay_nearest_N2} an example where
synchronization rate of the fastest mode is different for two different delay distributions with the same mean $\bar{\tau}$.

\begin{figure}[tb]
  \begin{center}
    \includegraphics[width=8.0cm]{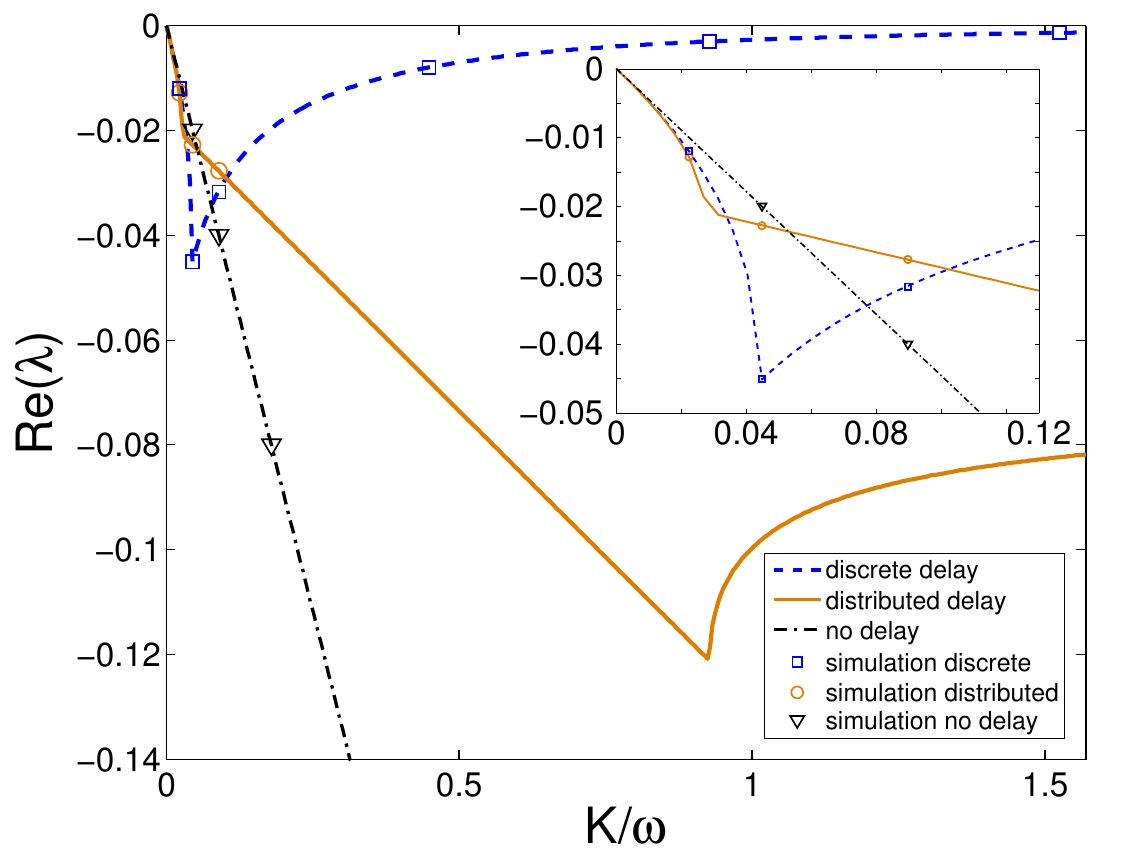} 
    \caption{Delayed coupling induces non-monotonic synchronization rate as a function of coupling strength. 
    $\Re\left( \lambda \right)$ versus $K/\omega$ for the eigenvalue $\lambda$ with the largest real part.
    Dashed curve: discrete delay. Continuous curve: distributed delay with  an exponential distribution.
    $N=2$, $n=1$, $\omega=0.223$~{min}$^{-1}$, $h(\theta)=\sin(\theta)$, $\bar{\tau}=2\pi/\omega\approx  28.176$~{min}.
    Dashed-dotted curve: no delay, $\bar{\tau}=0$. Symbols indicate values obtained from synchronization rates of numerical simulations.
    Inset: zoom for small $K/\omega$.}
    \label{fig:ReLambdaVsK}
  \end{center}
\end{figure}

%
To illustrate the effect of delays on synchronization dynamics, we calculate the synchronization rate for an exactly solvable example. 
We choose the simple case of $N=2$ mutually coupled oscillators, for which $n=1$, with sinusoidal coupling $h(\theta)=\sin(\theta)$. 
For this system the eigenvalues of the connectivity matrix $d$ are $\zeta_1=-1$ and $\zeta_2=1$.
We focus on the dependence of the synchronization rate with the coupling strength $K$, and choose to restrict the study to mean delay values such that $\Omega\bar{\tau}=2\pi$, 
for which the coefficient $\alpha$ is a constant equal to $K$ and the synchronized states given by Eq.~(\ref{eqn:collective-freq-coherent}) are always stable for positive $K$.
We study two extreme cases of delay distributions: discrete delay $g(s)=\delta(s-\bar{\tau})$ and an exponentially distributed delay,  $g(s)=\bar{\tau}^{-1}\,e^{-s/{\bar{\tau}}}$.
For discrete and distributed delays the non-trivial solution of the characteristic equation Eq.~(\ref{eqn:charact-eq-coherent-definitionzeta})
corresponding to the slowest decaying perturbation mode is found for $\zeta_1=-1$. 
For discrete delay this implies:		
\begin{equation}
  \lambda=-K+\frac{1}{\bar{\tau}}W\left(-{K\bar{\tau}}e^{K\bar{\tau}} \right),
  \label{eqnd:solutionLambertW}
\end{equation}
where $W(x)$ is the Lambert-W function \cite{Corless96}.
For the exponentially distributed delay the solution of the characteristic equation with $\zeta_1=-1$ is:
\begin{equation}
  \lambda=-\frac{1+ K \bar{\tau}}{2\bar{\tau}}\pm\frac{1}{2\bar{\tau}}\sqrt{(1+ K \bar{\tau})^2-8 K \bar{\tau}}.     
  \label{eqnd:K-depend-dist}
\end{equation}
In Fig.~\ref{fig:ReLambdaVsK} we plot the largest non-trivial $\Re(\lambda)$ as a function of $K$ for both kinds of delay.
Interestingly, there exists an optimal coupling strength for which the synchronization rate is maximal.
The optimal coupling strength depends on the shape of the delay distribution and is in stark contrast with non-delayed coupling, 
where stronger coupling strength always implies faster synchronization, Fig.~\ref{fig:ReLambdaVsK}.
The figure also shows that for weak coupling, the presence of a delay speeds up synchronization.

Not surprisingly, for $K\to 0$ the synchronization rate tends to zero for both kinds of delays. However, for discrete delays the same happens asymptotically as $K\to \infty$: 
synchronization becomes increasingly slower as coupling strength increases, and asymptotically the synchronized state is only neutrally stable.
In contrast, for distributed delays synchronization is robust, $\Re(\lambda)=-2/\bar{\tau}$ as $K\to \infty$.\\


\noindent
\emph{Discussion.}
We studied a system of oscillators coupled with distributed delays. 
We have shown that the collective frequency and the linear stability of the fully synchronized state
given by the solutions of Eq.~(\ref{eqn:collective-freq-coherent}) depend only on the mean of the delay distribution, and are independent of its shape. 
This suggests that discrete delays provide a good description of synchronized states, even if delays are distributed.
Close to the synchronized state, we found that transient dynamics depends on the shape of the delay distribution.
We have shown that in the presence of coupling delays, there can be a value of coupling strength that maximizes synchronization rate.
The observed optimal coupling depends on the shape of the delay distribution. 
Non-monotonic synchronization has been previously observed in models with phase shifts in the coupling~\cite{omelchenko12}, and
is similar to the enhancement of neural synchrony by coupling delays~\cite{dhamala2004}, also reported for other models~\cite{shrii2012}.
We have presented an example with $\Omega\bar{\tau}=2\pi$ where the synchronization rate vanishes for discrete delays and large coupling strength $K$, 
while it remains finite for distributed delays for any value of the coupling strength.
Furthermore,
for most values of $K$ the distributed delay gives faster synchronization than the discrete delay.
Altogether, this implies better overall robustness of the synchronization process with distributed delays in the coupling when compared to discrete delays.
However, note that there is a small range of coupling strength $K$ for $0.03 \lesssim K/\omega \lesssim 0.1$ in which discrete delay yields faster synchronization than distributed delays.
This might be a biologically relevant regime, since it is of the same order as $K/\omega=0.3$, the experimental estimation for the zebrafish segmentation clock coupling strength~\cite{Herrgen2010}.

The results shown in Fig. \ref{fig:ReLambdaVsK} can be tested experimentally, for instance using
electronic circuits of coupled phase-locked loops (PLL). These circuits provide well controlled conditions to study synchronization \cite{best2003,goldman2007,Pollakis2014,Jorg2015}. 
A setting with PLLs connected in parallel through elements introducing tunable delays would allow a test of our predictions. 
Also, cells exhibiting genetic oscillations and coupled via intercellular signaling pathways can be manipulated to change coupling strength and delays \cite{Herrgen2010,RiedelKruse2007}. 
The dependence of synchronization rate on coupling strength for different delay distributions can provide insight on the shape of the underlying delay distribution. 
Determination of the shape of delay distributions can be a source of information about the dynamics of molecular processes underlying signal transmission between cells.
This example shows the relevance of the study of synchronization rates for biological systems.

%
We thank David J. J\"{o}rg, Douglas B. Staple and the Oates Lab members for providing valuable comments. 
We acknowledge discussion with Gerhard Fettweis, Wolfgang Rave and Alexandros Pollakis.
We acknowledge the cfAED Cluster of Excellence of the TU Dresden.
S.A. acknowledges funding from
the Spanish Ministry of Economy and Competitiveness (MINECO)  through the Ram\'on y Cajal program.
%
%
%
L.G.M. and A.C.O. were supported by the Max Planck Society and the European Research Council under the European Communities Seventh Framework Programme (FP7/ 2007-2013)/ERC Grant No. 207634.
A.C.O. is supported by the Wellcome Trust. 


%

\end{document}